\documentclass{jfm}

\usepackage{graphicx}
\usepackage{natbib}
        
\ifCUPmtlplainloaded \else
  \checkfont{eurm10}
  \iffontfound
    \IfFileExists{upmath.sty}
      {\typeout{^^JFound AMS Euler Roman fonts on the system,
                   using the 'upmath' package.^^J}%
       \usepackage{upmath}}
      {\typeout{^^JFound AMS Euler Roman fonts on the system, but you
                   dont seem to have the}%
       \typeout{'upmath' package installed. JFM.cls can take advantage
                 of these fonts,^^Jif you use 'upmath' package.^^J}%
      }
  \else
  \fi
\fi

\ifCUPmtlplainloaded \else
  \checkfont{msam10}
  \iffontfound
    \IfFileExists{amssymb.sty}
      {\typeout{^^JFound AMS Symbol fonts on the system, using the
                'amssymb' package.^^J}%
       \usepackage{amssymb}%
         
       \let\ge=\geqslant  
      }{}
  \fi
\fi

\ifCUPmtlplainloaded \else
  \IfFileExists{amsbsy.sty}
    {\typeout{^^JFound the 'amsbsy' package on the system, using it.^^J}%
     \usepackage{amsbsy}}
    {}
\fi

\newsavebox{\astrutbox}
\sbox{\astrutbox}{\rule[-5pt]{0pt}{20pt}}

\usepackage{amsmath}
\usepackage{multirow}

\title{On the Kolmogorov Constants for the Second-Order Structure Function and the Energy Spectrum}

\author[R. Ni and K.-Q. Xia]
{R\ls U\ls I\ns  N\ls I\ns \and K\ls E -Q\ls I\ls N\ls G\ns X\ls I\ls A%
  \thanks{Email address for correspondence: kxia@phy.cuhk.edu.hk}}

\affiliation{Department of Physics, The Chinese University of Hong Kong, Shatin, Hong Kong, China}

\begin{document}
\maketitle

\begin{abstract}

We examine the behavior of the Kolmogorov constants $C_2$, $C_{k}$, and $C_{k1}$, which are, respectively, the prefactors of the second order longitudinal structure function, the three dimensional and one-dimensional longitudinal energy spectrum in the inertial range. We show that their ratios, $C_2/C_{k1}$ and $C_{k}/C_{k1}$, exhibit clear dependence on the micro-scale Reynolds number $R_{\lambda}$, implying that they cannot all be independent of $R_{\lambda}$. In particular, it is found that $(C_{k1}/C_{2}-0.25) = 1.95R_{\lambda}^{-0.68}$. The study further reveals that the widely-used relation $C_2 = 4.02 C_{k1}$ holds only asymptotically when  $R_{\lambda} \gtrsim 10^{5}$. It is also found that $C_2$ has much stronger $R_{\lambda}$-dependence than either $C_{k}$,  or $C_{k1}$ if the latter indeed has a systematic dependence on $R_{\lambda}$. We further show that  the variable dependence on $R_{\lambda}$ of these three numbers can be attributed to the difference of the inertial range in real- and wavenumber-space, with inertial range in real-space known to be much shorter than that in wavenumber space.

\end{abstract}


\section{Introduction}

The idea of homogeneous and isotropic turbulence (HIT) \citep{K41} enables one to focus on the essential physics of small-scale turbulent properties in the simplest possible case and use it as a first step to understand more complicated turbulence problems The Kolmogorov 1941 model (K41) \citep{K41} for HIT provides predictions of turbulence properties that agree well with experiments when the effect of intermittency is negligible.  One example is the ``two-thirds law" for second-order velocity structure functions and its counterpart in wavenumber space, the ``five-thirds law" for energy spectra, all expected to hold in the so-called  {\it inertial range}, i.e. 
\begin{equation}
\begin{split}
D_{LL}(r) = \langle\{[\bold{u}(\bold{x}+\bold{r})-\bold{u}(\bold{x})]\cdot\bold{\hat{r}}\}^2\rangle = C_2\epsilon^{2/3}{r}^{2/3},\\ E(k) = C_k\epsilon^{2/3}k^{-5/3}, ~~~~ E_{11}(k) = C_{k1}\epsilon^{2/3}k^{-5/3},
\label{eq:inert}
\end{split}
\end{equation}
here $D_{LL}$ is the second order longitudinal structure function, $E(k)$ is the three dimensional velocity spectrum, and $E_{11}(k)$ is the one dimensional longitudinal velocity spectrum, and $k$ is the wavenumber. 
The prefactors $C_2, C_k$ and $C_{k1}$ in front of the respective power laws are the so-called Kolmogorov constants. These constants are widely considered universal, i.e. independent of the flow field and the Reynolds number. Once these constants are known, they can be used to obtain the energy dissipation rate from the measured structure functions or spectra.

One of the most important parameters in any turbulent flows is the kinetic energy dissipation rate. A traditional method for measuring the energy dissipation rate is using  hot-wire probes while invoking Taylor's frozen flow hypothesis. However, Taylor's hypothesis requires that the mean velocity is much larger than the velocity fluctuation. Therefore, in turbulent systems with zero-mean velocity or wall-bounded flow with small mean velocity, single point velocity measurement could no longer be used to determine the dissipation rate. There is a growing consensus recently that the multipoint velocity measurement methods, such as particle image velocimetry (PIV) and particle tracking velocimetry (PTV), could be used to determine the energy dissipation rate both {\itshape{directly}} and {\itshape{indirectly}}. The direct method calculates the square of spatial derivatives of velocity from the original definition, which is more vulnerable to random errors. 
The indirect method for dissipation rate estimation makes use of the measured structure functions or velocity spectra in the inertial range via (\ref{eq:inert}). For high Reynolds number turbulence the indirect method has great advantages, since the inertial range covers the intermediate scales that are easier to measure. Therefore, many experiments with PIV or PTV velocity measurements used this method to determine the energy dissipation rate \citep{2008JFMSalazar}.
As is seen from (\ref{eq:inert}), in addition to the value of the measured $D_{LL}(r)$, the accuracy of the energy dissipation rate determined from this method also depends on the accuracy of $C_2$ used. The direct measurements on $C_2$ is very limited. For example, \citet{1987ZPBEffinger} compared theoretically obtained $C_2=1.7$ with those achieved experimentally $C_2=2.3$ \citep{1984JFMAnselmet, 1970JFMVanAtta}, and argued the difference is due to overestimation of dissipation rate in experiments. In many recent studies \citep{2008JFMSalazar, 2002JFMVoth}, the values of $C_{2}$ or $C_{k}$ are often obtained from that of $C_{k1}$ using \citep{my1975}
\begin{equation}\label{eq:c}
C_2 \approx 4.02 C_{k1},  ~~~~~~ C_{k} = \frac{55}{18}C_{k1},
\end{equation}
This is because the one-dimensional spectra is more accessible  experimentally and there exists a large set of data for $C_{k1}$ \citep{1995POFSreenivasan}. From (\ref{eq:c}) and $C_{k1}=0.53\pm0.055$ \citep{1995POFSreenivasan}, one may easily obtain that $C_2=2.13\pm0.22$, which is the most widely used value for $C_2$. A brief survey of the literature shows that ({\ref{eq:c}}) are very often used for finite values of $R_{\lambda}$ when in fact is derived under the idealized condition of the integral length $L=\infty$ and the Kolmogorov dissipation scale $\eta=0$ \citep{my1975}, which implies that equation~({\ref{eq:c}}) holds only for an infinite inertial range that may be realized only at infinite $R_{\lambda}$. As shown clearly in ({\ref{eq:inert}}), $C_{2}$, $C_{k}$ and $C_{k1}$ are not the coefficients of the second-order structure function and spectrum for the whole range of scales, but simply the prefactors of their power-law expressions valid {\it only in the inertial range}. 

In this paper, we examine the relationships among the three Kolmogorov constants from the experimental results in different turbulent systems and present two methods of determining the $R_{\lambda}$-dependent relationships between $C_{2}$ ($C_{k}$) and $C_{k1}$. The rest of this paper is organized as follows. We will first describe the experimental setups in our turbulent convection system as well as in other systems in \S \ref{sec:exp}. The difference between the experimental results with previous findings will be discussed in details in \S \ref{sec:SF2}. In \S \ref{sec:IR}, we will introduce a method to estimate the limits of finite inertial range and the $C_k$ from the second-order structure function. We will extend this method to the estimation of the $C_{k1}$ and the ratios among three Kolmogorov constants in \S \ref{sec:ck1}. In addition, another method based on the measured inertial range width will be discussed in \S \ref{sec:m2}. We will discuss about our findings and conclude in \S \ref{sec:con}.

\section{Experimental Setup}
\label{sec:exp}
Four sets of data were used in the present study and are labeled as Sets I, II, III and IV with their $R_{\lambda}$ listed in Table I. Set I comes from Rayleigh-B\'enard turbulent convection (RBC). Set II comes from an axisymmetric jet experiment \citep{1984JFMAnselmet}. The hot-wire for these measurements were positioned at a downstream distance $x/d=25$ from the jet nozzle of diameter $d=12$ cm at micro-scale Reynolds number $R_{\lambda}=536$, and $x/d=35$ at $R_{\lambda}=852$. Set III is from wind tunnel experiments at NASA Ames Research Center \citep{1994JFMSaddoughi}. The hot wires located on the centerline of the tunnel ceiling were used to measure the velocity. Set IV is from a direct numerical simulation of homogeneous and isotropic turbulence with the resolution ranging from $128^3$ to $1024^3$ grid points \citep{2002POFGotoh}. 

In the present experiment Lagrangian particle tracking velocimetry (PTV) was used to acquire data Set I. It was carried out in a convection cell using water as working fluid. The cell is a vertical cylinder with both its height and diameter being 19.2 cm, its top and bottom plates are made of copper and sidewall made of Plexiglas \cite{2005PRESun, 2012JFMNi}. The mean temperature of the bulk fluid is maintained at approximately $40^\circ$C, so the Prandtl number $Pr=\nu/\kappa=4.4$ ($\nu$ and $\kappa$ being kinematic viscosity and thermal diffusivity, respectively). The tracking volume is roughly $(5~cm)^3$ in the center of the cell, which is illuminated by an expanded laser beam. Three fast cameras were used to capture the images of seeding particles with diameter 50 $\mu$m and density 1.03 $g/cm^3$, and the acquired images were used as input in a computer program to reconstruct the 3D positions of those particles. The Stokes number of the particles in the experiment ranges from $10^{-3}$ to $10^{-4}$, indicating that the particles may be taken as tracers. The resolution for the particle positions is $\sim$8 $\mu$m, and the minimum resolvable separation between one pair of particles is typically $100-200~\mu$m because of the finite particle size and diffraction effect. The camera frame rates are 100 fps for all acquired data in Set I, which is sufficient to resolve dissipative range properties for the local Kolmogorov time scale $\tau_{\eta}=(\nu/\epsilon)^{1/2}$, with $\epsilon$ being the energy dissipation rate. The velocities were obtained by taking direct differentiation of the trajectories, which have been smoothened by a Gaussian kernel \citep{2002JFMVoth}. Other details of the setup and the PTV apparatus are described elsewhere \citep{2012JFMNi}.

\section{Results and discussions}

\subsection{Second-order structure function}
\label{sec:SF2}
\begin{figure}
\begin{center}
$\begin{array}{cc}
\includegraphics[width=3.4in]{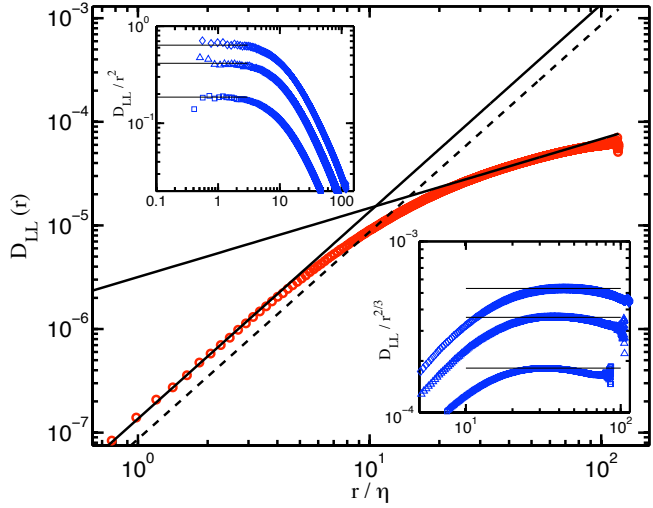} 
\end{array}$
\caption{A second-order longitudinal structure function ($R_{\lambda} = 89.7$). Solid lines: dissipative range and inertial range limit. Dashed line: see text. Upper and lower insets: $D_{LL}(r)$ compensated by $r^2$ and $r^{2/3}$, respectively. The symbols, from top to bottom, correspond to $R_{\lambda}=$ 89.7, 71.8, and 55.0 (Set I). The solid lines are fits to the data in the respective plateaus from which $\epsilon$ and $C_2$ are obtained.}
\label{fig:dissrange}
\end{center}
\end{figure}
Set I was measured in the central region of a cylindrical RBC cell \citep{2012JFMNi}, in which the velocity field has been found to be approximately homogeneous and isotropic, with the structure functions in the inertial range the same as those found for homogeneous and isotropic turbulence \citep{2006PRLSun,2008JFMZhou,2010ARFMXia}. Figure~{\ref{fig:dissrange}} shows an example of $D_{LL}(r)$ from this Set. The upper inset plots $D_{LL}$ for several values of $R_{\lambda}$, compensated by the dissipative range scaling $r^2$, and the lower inset plots $D_{LL}$ compensated by the the inertial range scaling $r^{2/3}$. The solid lines in the upper inset give the plateau height averaged over data points on the plateau. Using $D_{LL}(r) = (\epsilon/15\nu)  r^2$ for the dissipative range ($r\ll\eta$), we obtain the energy dissipation rate $\epsilon$ \citep{2011PRLNi}. It is seen in the lower inset that the plateau that may exist, if at all, must be very short. This is due to the limited Reynolds number. Still, we could approximate ({\ref{eq:inert}}) to obtain $C_2$ by only taking the average over data points near the peak as shown by those solid lines. For the data with $R_{\lambda} =89.7$ in figure~{\ref{fig:dissrange}}, this method gives $C_2 = 1.56$, which is much lower than the value of 2.13 given by ({\ref{eq:c}}) with $C_{k1} = 0.53$ \citep{1995POFSreenivasan}. If we use $C_2 = 2.13$ to estimate $\epsilon$ via ({\ref{eq:inert}}), the obtained value would be $37\%$ smaller than that directly measured from dissipative range second-order structure function. This is illustrated by the dashed line in figure~{\ref{fig:dissrange}}, which indicates $D_{LL}(r)$ in the dissipative range according to $\epsilon$ based on $C_2 = 2.13$.

\subsection{Method I}
\label{sec:IR}
We now present the first method of determining the $R_{\lambda}$-dependent relationships among the Kolmogorov constants from measured structure functions. We note that ({\ref{eq:c}}) were derived from the integral transforms between the second-order structure function and energy spectra, which, in dimensionless form, are written as \citep{my1975}
\begin{equation}
\begin{split}
\beta_{LL}(x) = 4\int^{\infty}_{0}\left[ \frac{1}{3}+\frac{cos\xi x}{(\xi x)^2}-\frac{sin\xi x}{(\xi x)^3}\right] \varphi (\xi) d\xi, \\
\label{eq:beta}
\end{split}
\end{equation}
\begin{equation}
\varphi_{11}(\xi) = \int^{\infty}_{\xi}\left( 1-\frac{\xi^2}{\xi'^2}\right)\frac{\varphi(\xi')}{\xi'}d\xi',
\label{eq:threetoone}
\end{equation}
where $\beta_{LL}(r/\eta)= D_{LL}(r) / u_{\eta}^2$, $\varphi(\eta k) =E(k)/(\eta u_{\eta}^2)$, and $\varphi_{11}(\eta k) = E_{11}(k) / (\eta u_{\eta}^2)$, with $u_{\eta} = (\nu\epsilon)^{1/4}$ the Kolmogorov velocity scale; and $x=r/\eta$, $\xi = k\eta$. One can break up the integral
in (\ref{eq:beta}) into three parts with respective limits: $[0,\xi_1)$, $[\xi_1,\xi_2]$ and $(\xi_2,\infty)$, where $\xi_1$ and $\xi_2$ are the yet to be-defined boundaries of the inertial range. The three parts correspond to integrations over large-scale, inertial range, and dissipative range, respectively. In the idealized case of infinite inertial range, i.e. $\xi_1\rightarrow 0$ and $\xi_2\rightarrow \infty$, the integrals over the dissipative range and the large-scale vanish and one readily obtains ({\ref{eq:c}}) by substituting the inertial range results $\beta_{LL}(x) = C_2x^{2/3}$, $\varphi(\xi) = C_k\xi^{-5/3}$ and $\varphi_{11}(\xi) = C_{k1}\xi^{-5/3}$  into ({\ref{eq:beta}}) and ({\ref{eq:threetoone}}). This proves our assertion earlier that ({\ref{eq:c}}) are valid only for infinite inertial range, corresponding to infinite $R_{\lambda}$.

\begin{figure}
\begin{center}
\includegraphics[width=3.1in]{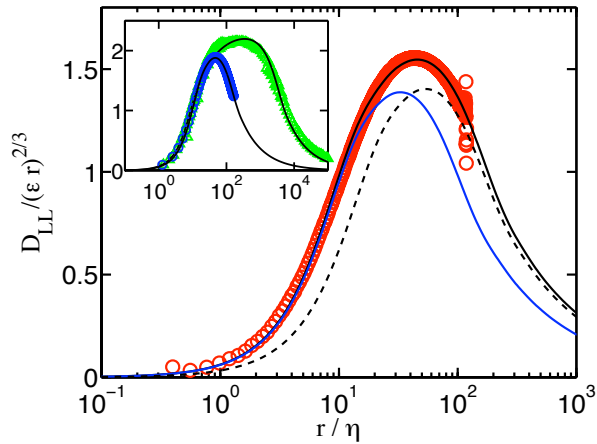}
\caption{(colour online)  Circles: compensated $D_{LL}$ for $R_{\lambda}$ = 89.7 (Set I).
Black line: A fit of ({\ref{eq:beta}}) to the measured structure function. Dashed line and blue line: see text. Inset:  More examples of compensated $D_{LL}$ with fittings. Blue circles: $R_{\lambda} = 70$ (Set IV); green triangles: $R_{\lambda}=600$ (Set III).}
\label{fig:fitpara}
\end{center}
\end{figure}

To determine $\xi_1$ and $\xi_2$ for finite $R_{\lambda}$, we note that the integral over $[0,\xi_1)$ is small because  the function in the bracket before $\varphi (\xi)$ in ({\ref{eq:beta}}) approaches zero for very small $\xi$ \citep{my1975}. The integral over $(\xi_2,\infty)$  is also negligible because the energy spectrum in dissipative range decays to zero exponentially \citep{1948PRSLAHeisenberg}. Thus, to a good approximation one may drop these two integrals. The three parameters $\xi_1$, $\xi_2$ and $C_k$ can then be determined by fitting the remaining integral over inertial range (with $\varphi(\xi) = C_k\xi^{-5/3}$) to the measured second-order structure function. The red circles in figure~{\ref{fig:fitpara}} shows $D_{LL}/(r\epsilon)^{2/3}$  ($R_{\lambda} = 89.7$, Set I) with the solid line as fitting result, which is seen to be excellent. Table I lists the fitted parameters $\xi_1$, $\xi_2$ and $C_{k}$ for all data sets.  To check whether the fitting result is unique, we changed the fitted value of  $\xi_2$ to $0.6\xi_2$ while keeping the other two parameters the same, and used these to evaluate the integral in ({\ref{eq:beta}}), which produced the dashed line in the figure. It is seen that the right tail of this line is very close to the original fitting line while its left tail shifted from the data, suggesting that $\xi_2$ controls the transition from dissipative scale to inertial range. Similarly, we changed $\xi_1$ to $1.6\xi_1$, which gave the blue solid line in the figure. It shows that the left tail remains unchanged while the right tail shifted, suggesting that $\xi_1$ controls the transition from the inertial range to the large-scale. It is also obvious that  $C_k$ controls the plateau value of the compensated second-order structure function in the inertial range. Evidently, the measured second-order structure function has three different regions: transition zone from dissipative to inertial range, the inertial range, and transition zone to large-scale. Thus, for a given compensated second-order structure function with left and right tails and a plateau (or a peak), once $\xi_1$ and $\xi_2$ are fixed, the value of $C_{k}$ is uniquely determined. In the inset of figure~{\ref{fig:fitpara}} we show two more examples, which again shows good fitting results.

\begin{figure}
\begin{center}
\includegraphics[width=3.1in]{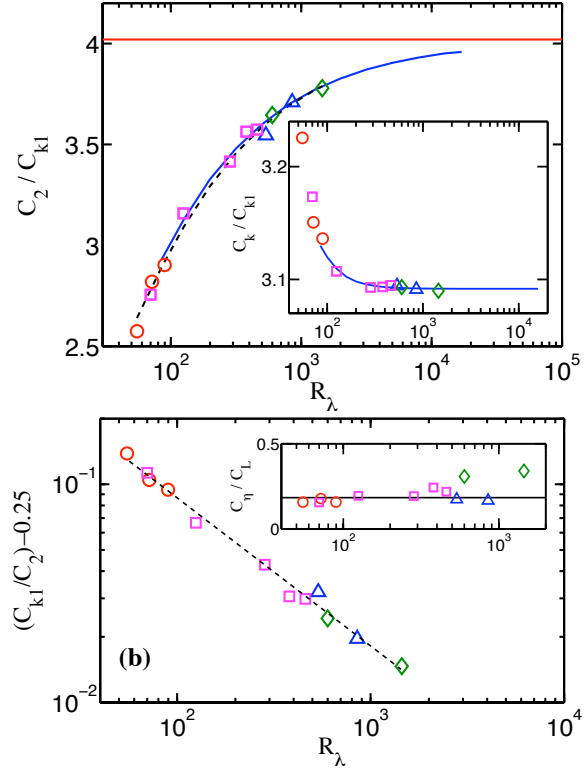}
\caption{(Colour online) (a) Symbols in the main figure and the inset: ratio of $C_2/C_{k1}$ and $C_{k} / C_{k1}$ determined from the four data sets with $C_{2}$ directly from second-order structure functions, and  $C_{k}$ and $C_{k1}$ from fitting (\ref{eq:beta}) and (\ref{eq:threetoone}) to second-order structure functions. Dashed and solid lines: see text. The horizontal red line has a value of 4.02. Throughout this figure, the symbols are: circles: Set I; triangles: Set II; diamonds: Set III; squares: Set IV. (b) Symbols same as in (a) but are plotted as  $(C_{k1}/C_{2})-0.25$ vs. $R_{\lambda}$. Dashed line is also the same as in (a). Inset: ratio of the parameters $a_{\eta}$ and $a_{L}$ that characterize the inertial range. Solid line: average value of all points excluding diamonds.}
\label{fig:cover}
\end{center}
\end{figure}

\begin{figure}
\begin{center}
$\begin{array}{cc}
\includegraphics[width=3in]{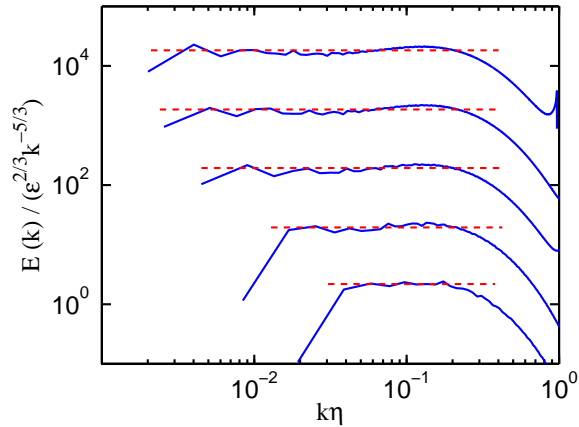} 
\end{array}$
\caption{(Colour online) Compensated three dimensional spectra taken from Set IV \citep{2002POFGotoh}. From top to bottom, $R_{\lambda}=$460, 380, 284, 125, and 70. The red dashed lines are based on parameters from the structure-function-fitting method: they span horizontally from $\xi_1$ to $\xi_2$ with height equal to $C_k$. The spectra are shifted vertically for clarity.}
\label{fig:spectrum}
\end{center}
\end{figure}

\begin{table}
\begin{center}
\begin{tabular}{c c c c c c c c c}
$Source$ & $R_{\lambda}$ & $C_2$ & $\xi_1$  & $\xi_2$ &$C_k$ & $C_{k1}$ & $a_{L}$ & $a_{\eta}$\\
 I& 55.04 & 1.210 & 0.0515 & 0.48 & 1.516 & 0.470 & 3.06 & 0.48 \\ 
& 71.79 & 1.404 & 0.0332 & 0.49 & 1.569 & 0.498 & 2.76 & 0.49 \\ 
  & 89.72 & 1.556 & 0.0278 & 0.48 & 1.681 & 0.536 & 3.06 & 0.48 \\ 
II& 536 \footnote{This point has a large fitting error due to the limited amount of data points in the structure function.}& 2.067 &0.0037 & 0.68 & 1.804 & 0.583 & 3.93 &  0.68 \\ 
 & 852 & 2.400 & 0.0019 & 0.58 &  2.000 & 0.647 & 3.51 & 0.58\\
III & 600 & 2.151 & 0.0013 & 0.49 & 1.825 & 0.590 & 1.59 & 0.49 \\
 & 1450 & 2.184 & 0.0004 & 0.51& 1.786 & 0.578 & 1.50 & 0.51 \\
IV& 70 & 1.893 & 0.0305 & 0.38 & 2.180 & 0.687 & 2.45 & 0.38 \\
 & 125 & 1.974 & 0.0129 & 0.42 & 1.942 & 0.625 & 2.16 & 0.42 \\
 & 284 & 2.131 & 0.0045 & 0.41 & 1.930 & 0.624 & 2.14 & 0.41 \\
 & 380 & 2.138 & 0.0024 & 0.40 & 1.856 & 0.600 & 1.65 & 0.40 \\
 & 460 & 2.112 & 0.0021 & 0.40 & 1.829 & 0.591 & 1.83 & 0.40 \\
\end{tabular}
\end{center}
\caption{Parameters and fitting results for the four  data sets used in the present work. Set I: from this work; Set II: axisymmetric jet \citep{1984JFMAnselmet}; Set III: wind tunnel experiments \citep{1994JFMSaddoughi}; Set IV: direct numerical simulation \citep{2002POFGotoh}.}
\label{tab:para}
\end{table}

\subsection{Relations among different Kolmogorov constants}
\label{sec:ck1}
Although the relationships between $C_2$ with $C_k$ can be established with a given $D_{LL}$,  it is the one-dimensional spectrum that is more experimentally accessible. So it is of more practical importance to find relationships between $C_{k1}$ and $C_k$ ($C_2$). For the integral in ({\ref{eq:threetoone}),  we take the same treatment for the inertial range and dissipative scales as above, i.e. $\varphi(\xi) = C_k\xi^{-5/3}$ in $[\xi_1,\xi_2]$ and $0$ in $(\xi_2,\infty)$. As we are interested in the inertial range, we restrict the value of $\xi$ in ({\ref{eq:threetoone}) to $\xi \ge \xi_1$ so we do not need to concern the form of $\varphi(\xi)$ for $\xi < \xi_1$ [so the obtained spectrum $\varphi_{11}(\xi)$ is valid only in the inertial range].
With this procedure, we obtain $C_{k1}$  by integrating ({\ref{eq:threetoone}) with the fitted  $\xi_1$, $\xi_2$ and $C_k$, and the results are listed in Table I. The values of either $C_{2}$, or $C_{k}$ or $C_{k1}$ in the table reveal no clear trend with $R_{\lambda}$ from one set of data to the next or within one set. From ({\ref{eq:inert}}), we see that the absolute values of $C_{2}$ and $C_{k}$ ($C_{k1}$) determined as the plateaus of the compensated structure function/spectra will be affected by the errors in $\epsilon$, but such errors will be canceled out if we take their ratios. As shown by the symbols in figure~{\ref{fig:cover}}(a) and the inset, this appears to be the case, i.e. clear trends emerge with $R_{\lambda}$ for these ratios. To find an empirical relation between $C_{2}$ and $C_{k1}$, we fit the symbols in figure~{\ref{fig:cover}}(a) with $C_{2}/C_{k1} = 1/(0.25+AR_{\lambda}^{\beta})$. The result is shown as the dashed-line in the figure, with the fitting parameters $A = 1.95$ and $\beta = -0.68$. In figure~{\ref{fig:cover}}(b) we plot  the data as $(C_{k1}/C_{2})-0.25$ vs. $R_{\lambda}$ with  the dashed-line representing the fitting, which shows that the plotted quantity indeed follows a power-law within the data range, suggesting the fitting curve in figure~{\ref{fig:cover}}(a) can describe well the behavior of $C_{2}/C_{k1}$.

\subsection{Method II}
\label{sec:m2}
We now introduce a second method, based on experimental values of $L/\eta$, to determine the ratios of $C_{2}/C_{k1}$ and $C_{k}/C_{k1}$, but not the individual constants. Let  $\xi_1=a_{L}(\eta/L)$ and $\xi_2=a_{\eta}$, respectively. Table I lists the values of $a_{L}$ and $a_{\eta}$ for the four sets of data. It is clear that these numbers depend on $R_{\lambda}$, but their ratio appears to be independent of $R_{\lambda}$  [see the inset of figure~{\ref{fig:cover}}(b)] (the diamonds are from Set III and their large deviations from the others may be due to the noise in the original data \citep{1994JFMSaddoughi}. The horizontal line in the figure represents $a_{\eta}/a_{L} = 0.183$ and is an average based on Sets I, II and IV. [If all data points are included, the value will be 0.21 instead. But the difference in the obtained $C_{2}/C_{k1}$ in using either one of these values is small.]  It thus appears that although both the inertial range itself (characterized by $L/\eta$) and its start and end points (represented by $\xi_{1}$ and $\xi_{2}$ in $k$-space) depend on $R_{\lambda}$, the ratio $a_{\eta}/a_{L} = (\xi_{2}/\xi_{1})/(L/\eta)$ seems to be universal.  We note that if $\xi_{1}$ and $\xi_{2}$ are known for a given $R_{\lambda}$, we can then simply substitute (\ref{eq:inert}) into the integrals in (\ref{eq:beta}) and (\ref{eq:threetoone}) to obtained the ratios, without the need for either structure function or spectra data. For those data collected by \citet{2004PRECleve}, we obtain values of $L/\eta$, which are based on a number of experiments with $R_{\lambda}$ varying from 85 to 17,000. A  power-law fit to these gives $L/\eta =0.37R_{\lambda}^{1.27}$.  It is seen from Table 1 that $a_{\eta}$ is generally between 0.4 and 0.6. Taking $a_{\eta} = 0.5$, and with $a_{\eta}/a_{L} = 0.183$, we obtain $\xi_{2}$ and $\xi_{1}$ as functions of $R_{\lambda}$. For each of these pairs, we obtain $C_{2}/C_{k}$ and $C_{k}/C_{k1}$ via equations (\ref{eq:inert}), (\ref{eq:beta}) and (\ref{eq:threetoone}), and therefore $C_{2}/C_{k1}$. The results are shown as the solid lines in figure~{\ref{fig:cover}}(a) and the inset, respectively.  One sees there is excellent agreement with those obtained from fitting structure function data, as well as with the fitted curve.   We remark that if one takes $a_{\eta}$ to be any value between 0.4 and 0.6, the obtained $C_{2}/C_{k1}$ is essentially the same, especially when $R_{\lambda} > 100$. It should be note that in most real flows there exist no sharp boundaries between the three regions, i.e. the dissipative range, the inertial range and the large scales, but only transition zones. In this respect,  the parameters  $\xi_{1}$ and $\xi_{2}$ introduced here should be viewed as a mean to obtain the ratio of the Kolmogorov constants, which we have shown to be largely determined by the ratio $\xi_{2}/\xi_{1}$ and is insensitive to the specific values of these parameters if they are within a range.

As a self-consistent check of the structure-function-fitting method, we compare the obtained $\xi_{1}$, $\xi_{2}$ and $C_{k}$ with 3D spectra, which are from the same DNS study as the structure functions from Set IV but were obtained separately \citep{2002POFGotoh}. Figure~{\ref{fig:spectrum}} plots 5 compensated spectra with $R_{\lambda}$ from $70 - 460$. The horizontal dashed lines are determined by the three fitted parameters, i.e. they start  at $\xi_{1}$, end at $\xi_{2}$, and have heights equal to $C_{k}$. It is seen that the method can predict accurately not only the height of the spectrum plateau, but also the position and the width of the inertial range except the ending points, which are a bit larger than those shown in the spectra.

\section{Discussions and Conclusion}
\label{sec:con}

In the paper we have introduced two different methods that give the relationships among the three Kolmogorov constants $C_2$, $C_{k}$, and $C_{k1}$. The first method can be used to determine $C_k$ and $C_{k1}$ from the experimentally measured second-order structure functions by splitting the integral relation between the structure function and the energy spectrum into three different regimes: dissipative range, inertial range, and range for large scales. It is found that the ratio $C_2/C_{k1}$ exhibits clear dependence on the micro-scale Reynolds number $R_{\lambda}$. The ratio could be well fitted as $(C_{k1}/C_{2}-0.25) = 1.95R_{\lambda}^{-0.68}$. The second method directly determines the ratios among the three Kolmogorov constants from the experimentally-determined width of the inertial range without fitting to the second-order structure function. Our results reveal that the widely-used relation $C_2 = 4.02 C_{k1}$ holds only asymptotically when  $R_{\lambda} \gtrsim 10^{5}$. 

It is generally known that $C_{2}$, $C_{k}$ and $C_{k1}$ are asymptotic values, which will become a constant for high enough Reynolds number. Since the $C_{k1}$ is generally considered as constant for $R_{\lambda}>50$ and it is more experimentally available, $C_2$ are usually obtained from $C_2=4.02C_{k1}$ for $R_{\lambda}$ from hundreds to thousands. However, this is only valid if $C_2$ equals to $4.02C_{k1}$ for all $R_{\lambda}$. And of course, the $C_2$ should be constant for $R_{\lambda}>50$. In this work, we have shown that the $C_2/C_{k1}$ are Reynolds number dependence at $R_{\lambda}$ as high as $10^4$, which indicates that $C_2$ does not reach asymptote at least for $R_{\lambda}<10^4$. 

The reason for the stronger $R_{\lambda}$ dependency of $C_2$ is because there is no clear inertial range for second-order structure function for $R_{\lambda}$ even at $10^4$. It is found that if the structure function showed in log-log plot, there may have clear inertial range, however if the SF2 is compensated and plotted in log-linear scale, the inertial range may not be that apparent. It is consistent with previous findings that there is no inertial range for SF2 at $R_{\lambda}=19500$ with a more sensitive local slope test. In all $R_{\lambda}$ examined, we approximated ({\ref{eq:inert}}) to obtain $C_2$ by only taking the average over data points near the peak in the inertial-range-compensated structure function. The reason we could still use this is because, physically, the inertial range does exist for all $R_{\lambda}$ we used. There is an inertial range in three-dimensional energy spectrum which is the prerequisite for our method. But after transformation through (\ref{eq:beta}), the range is compressed in structure function form. And there are two benefits for this method. First, it could help us to determine when the $C_2$ will reach the asymptote. By extrapolation from our works, we estimate it is roughly at $R_{\lambda}=10^5$. Second, even with this approximated $C_2$, the experimentalist still could obtain the dissipation rate through the structure functions. The relations produced by our analysis would give a more accurate result than simply using $C_2=4.02C_{k1}$ for estimation. 

It is found before that both $C_2$ and $C_{k1}$ have great uncertainties such as those compiled by \citet{my1975} and by \citet{1995POFSreenivasan}. The uncertainties would come from the method in determining dissipation rate or the energy injection method in different turbulent systems. We may ascribe both ways to the error of energy dissipation rate. When comparing the Kolmogorov constants obtained in different systems, if there is a large uncertainty, any trend below would be hidden in the scatter exhibited by the data. Therefore, in this work, we consider the Reynolds number dependencies for the ratios rather than the value for each constant separately, since the contributions from energy dissipation error will be cancelled with each other.

We gratefully acknowledge support of this work by the Research Grants Council of Hong Kong SAR (No. CUHK404409 and N$\_$CUHK462/11).
\bibliographystyle{jfm}

\end{document}